\documentclass[prl, twocolumn, superscriptaddress]{revtex4}
\usepackage{graphicx,color}
\usepackage{amssymb}
\usepackage{amsmath}
\usepackage{bm}
\usepackage[colorlinks=true,linkcolor=blue,urlcolor=blue,citecolor=green]{hyperref}
\usepackage{tabularx}
\usepackage{multirow}
\usepackage{braket}
\usepackage{xcolor}
\usepackage{gensymb}
\usepackage{amsmath,amssymb}

\begin{document}

\title{Orbital magnetization as the origin of the nonlinear Hall effect}
\author{Zesheng Zhang}\thanks{These authors contributed equally to this work}
\affiliation{State Key Laboratory of Quantum Functional Materials, School of Physical Science and Technology, ShanghaiTech University, Shanghai 201210, China}
\author{Xin-Zhi Li}\thanks{These authors contributed equally to this work}
\affiliation{State Key Laboratory of Quantum Functional Materials, School of Physical Science and Technology, ShanghaiTech University, Shanghai 201210, China}
\author{Wen-Yu He}\thanks{hewy@shanghaitech.edu.cn}
\affiliation{State Key Laboratory of Quantum Functional Materials, School of Physical Science and Technology, ShanghaiTech University, Shanghai 201210, China}

\date{\today}

\begin{abstract}
The nonlinear Hall effect is a new type of Hall effect that has recently attracted significant attention. For the physical origin of the nonlinear Hall effect, while orbital magnetization has long been hypothesized to underpin the nonlinear Hall effect, a general relation between the two quantities remains elusive. Here, we resolve the problem by deriving the first explicit formula connecting the electric field induced orbital magnetization to the second order Hall conductivity. Our theory reveals that the applied electric field plays dual roles in generating the nonlinear Hall effect: it first generates nonequlibrium orbital magnetization associated with an edge current, and subsequently perturbs the circulating edge states to produce transverse Hall voltage. For the experimental verification, we propose to apply a combination of direct and alternative currents to identify the circulating edge current in the nonlinear Hall effect. Based on the orbital magnetization origin, we point out that in isotropic chiral metals of T and O point groups, the crystalline symmetry suppresses the nonlinear Hall response for a monochromatic linear polarized electric field, but a non-collinear bichromatic electric field can generate a finite nonlinear Hall current that manifests the chiral correlation of the field. This discovery finally enables us to incorporate both the nonlinear Hall effect and circular photo-galvanic effect into the framework of orbital magnetization.
\end{abstract}

\maketitle

\emph{Introduction}.--- The nonlinear Hall effect is a Hall phenomenon where a nonlinear voltage transverse to the applied electric field is generated in materials~\cite{Intisodeman, Qiongma, Fai01, Haizhou01, Ortix, Narayan, YuewenFang}. As a new member of the Hall effect family, the nonlinear Hall effect has sparked active investigations due to its broad application prospect and deep connection to quantum geometry~\cite{Qiongma, Fai01, Kumar, Ortix02, Yugui, Meizhen01, NingWang, ZhiqiangMAO, Zhimin01, Changgan, GaoAn, WeiboGao, WeiboGao2, Sankar, YangGao01, DiXiao, Shengyuan, Kalan, Haizhou2, YangZhang01, Isobe, Cano, Liangfu02}. Recent studies have ascribed the intrinsic mechanism of the nonlinear Hall effect to a series of quantum geometric quantities including Berry curvature dipole~\cite{Intisodeman, Qiongma, Fai01}, quantum metric dipole~\cite{YangGao01, DiXiao, Shengyuan, Kalan, GaoAn, WeiboGao}, and other gauge invariant terms~\cite{Chengping, Longjun, Agarwal, Wenyu01}. Those Berry curvature related quantum geometric quantities underlie many current studies of the nonlinear Hall effect.

Yet, a pivotal question persists: what is the microscopic origin of the nonlinear Hall effect? In the linear regime, the Berry curvature is intimately tied to both the Hall conductivity~\cite{Thouless, Nagaosa, QianNiu} and orbital magnetization~\cite{QianNiu, QianNiu02, Thonhauser, Ceresoli, Resta01}. It is known that the Berry curvature induced Hall conductivity originates from the edge current that simultaneously causes orbital magnetization~\cite{Halperin, Yoshioka}. Such intrinsic connection between the Hall conductivity and the orbital magnetization is speculated to also apply to the nonlinear Hall effect, where the second order Hall conductivity has been postulated to originate from the current induced orbital magnetization~\cite{Fai01, ZhiminLiao, Manchon, Kusunose}. Recent experimental observations in the few layer WTe$_2$~\cite{Qiongma, Fai01, Zhimin01, Ralph}, TaIrTe$_4$~\cite{Kumar, Zhimin03}, strained monolayer MoS$_2$~\cite{JieunLee, JieunLee2} and WSe$_2$~\cite{ZhiminLiao} show that the nonlinear Hall effect and the current induced orbital magnetization are companion phenomenon, which further supports the orbital magnetization as the origin of the nonlinear Hall effect. However, current understanding of the nonlinear Hall effect in terms of the orbital magnetization remains phenomenological, as the quantitative relation between the two effects has yet been established.

\begin{figure}[htbp]
    \centering
    \includegraphics[width=0.48\textwidth]{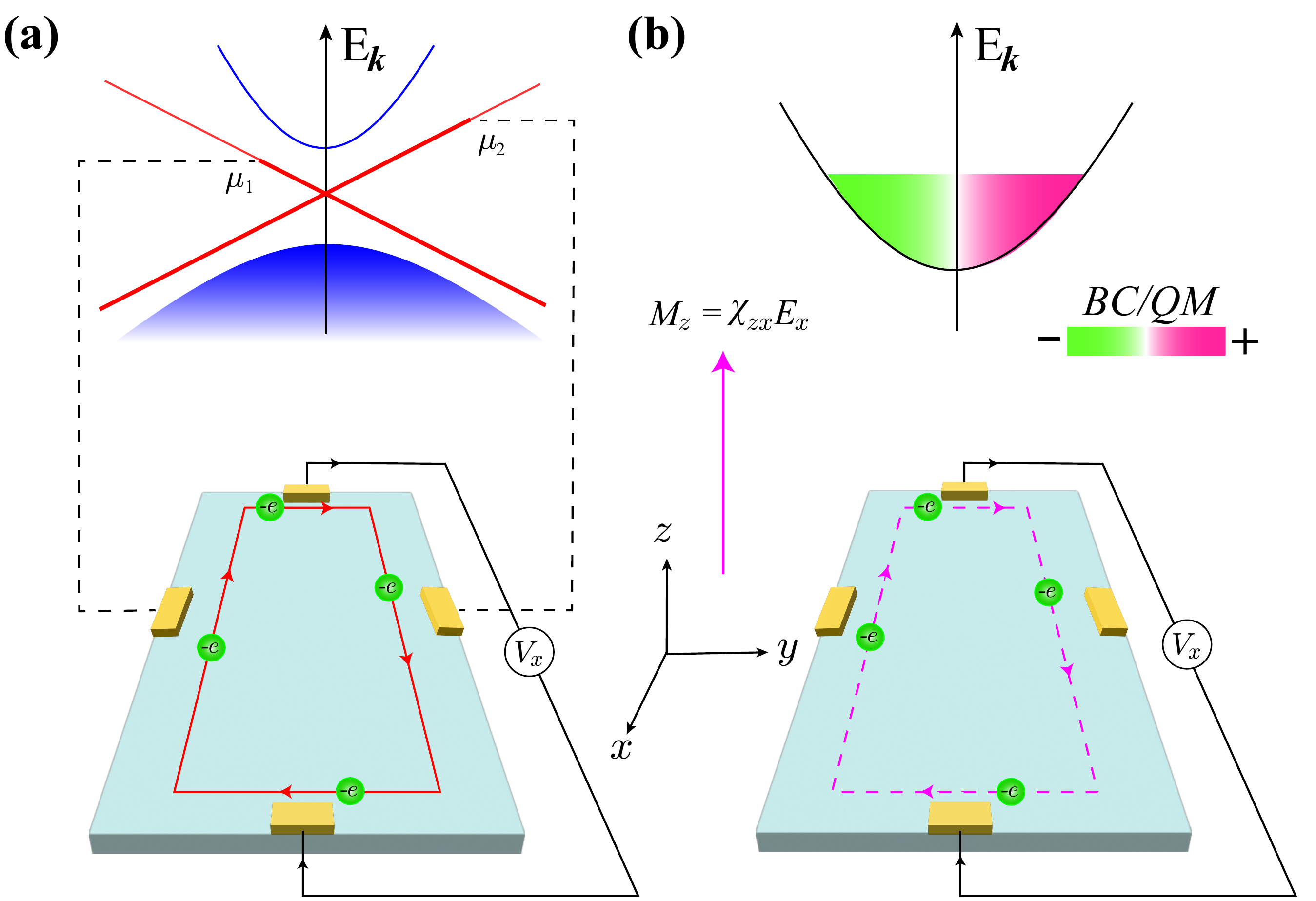}
    \caption{The Hall effect and the associated orbital magnetization. (a) In the linear Hall effect, a $z$-directional orbital magnetization $M_z$ arises from the circulating edge current (the red loop). An $x$-directonal electric field $E_x$ creates chemical potential difference at sample edges, yielding a Hall voltage in the $y$ direction. (b) The second order nonlinear Hall effect in metals with finite Berry curvature (BC) dipole and quantum metric (QM) dipole on Fermi surfaces. The first order effect of $E_x$ is to induce an orbital magnetization $M_z=\chi_{zx}E_x$, and the induced $M_z$ also has a corresponding edge current (the purple dashed loop). The second order effect of $E_x$ differs the chemical potentials at sample edges, generating the second order Hall voltage in the $y$ direction of the current loop.}
    \label{fig1}
\end{figure}

In this work, we resolve the problem by showing explicitly how the second order nonlinear Hall effect observed in metals arises from the electric field induced nonequilibrium orbital magnetization at the linear order. Past studies have demonstrated that the quadratic coupling of electric field components via dipoles of Berry curvature and quantum metric gives rise to second order nonlinear Hall current~\cite{Intisodeman, YangGao01, DiXiao, Shengyuan}. Here, we reveal that in generating the nonlinear Hall effect, the two electric field components in the quadratic coupling are individually involved in two successive processes: one component firstly creates net orbital magnetization associated with a circulating edge current, and the other one subsequently acts as a perturbation to induce the transverse Hall voltage (see Fig.\ref{fig1}). When the two successive processes are combined, the nonlinear Hall effect naturally occurs.

After clarifying the orbital magnetization as the physical origin of the nonlinear Hall effect, we propose that applying a combination of direct current (DC) and alternative current (AC) can experimentally verify the individual roles of the electric field components in generating the nonlinear Hall effect. Moreover, we get inspired to predict a unique nonlinear Hall effect driven by a non-collinear bichromatic electric field in isotropic chiral metals of T and O point groups. Here the non-collinear bichromatic electric field means that the non-collinear components of the field  have different frequencies. We show that the non-collinear components of the bichromatic electric field mutually act as perturbations to their induced orbital magnetizations, giving rise to the sum frequency generation~\cite{Boyd} in the direction perpendicular to the polarization plane of the non-collinear bichromatic electric field. Importantly, we find that the nonlinear Hall current in isotropic chiral metals directly manifests the chiral correlation of the applied electric field, which unifies the nonlinear Hall effect and circular photo-galvanic effect within the paradigm of electric field induced orbital magnetization. Finally, we propose a series of B20 transition metal monosilicides~\cite{Burkov} of T point group as the candidate materials that can exhibit the unique nonlinear Hall effect driven by the non-collinear bichromatic electric field.

\emph{Orbital magnetization and Hall effect}.--- The intrinsic connection between the Hall effect and orbital magnetization can be traced back to the classical scenario of Lorentz force induced electrons' cyclotron motion, which leads to the Hall effect and orbital magnetization simultaneously. In the classical picture, the cyclotron orbits of electrons bounce off the edge, forming chiral edge states circulating around the sample~\cite{Kane}. The picture of chiral edge states plays a central role in the modern theory of Hall effect. As the electronic chiral edge states circulate around the sample, the electrons' center-of-mass motion generates an orbital magnetization~\cite{QianNiu, QianNiu02, Thonhauser, Ceresoli, Resta01}
\begin{align}\label{OM_edge}
\bm{M}=&\frac{e}{\hbar}\sum_a\int_{\bm{k}}\bm{\Omega}_{a,\bm{k}}\frac{1}{\beta}\log\left[1+e^{-\beta\left(E_{a,\bm{k}}-\mu\right)}\right],
\end{align}
with $\beta^{-1}=k_{\textrm{b}}T$, $\int_{\bm{k}}\equiv\int d\bm{k}/\left(2\pi\right)^d$, $\bm{\Omega}_{a,\bm{k}}$ being the Berry curvature and $E_{a,\bm{k}}$ denoting the band dispersion. Here the subscript $a$ labels the band index and $d=2,3$ is the dimension. When an electric field is applied in the sample plane, the chemical potential $\mu$ across the sample becomes spatially inhomogeneous~\cite{Yoshioka}. According to the standard definition of orbital magnetization~\cite{Jackson}, the Hall current density at the linear order is derived to take the form
\begin{align}\label{J_perp}
J_{H,i}=\epsilon_{ijk}\frac{\partial M_{k}}{\partial r_j}=\epsilon_{ijk}\frac{\partial\mu}{\partial r_j}\frac{\partial M_{k}}{\partial\mu}.
\end{align}
Here we have focused on the macroscopic measurable Hall current and discounted the local orbital magnetic moments that only perturb the local current at the microscopic scale~\cite{Supp}. In the electric dipole approximation~\cite{Sipe}, the linear gradient of chemical potential is equal to the applied electric field : $\bm{E}=-\frac{1}{e}\partial_{\bm{r}}\mu$~\cite{Resta01,Yoshioka}, so Eq. \ref{J_perp} can be rearranged into the form of $J_{H,i}=\sigma^{H}_{ij}E_j$. Combining Eq. \ref{OM_edge} and Eq. \ref{J_perp}, we recover the Berry curvature induced linear Hall conductivity to be
\begin{align}\label{Hall01}
\sigma^{H}_{ij}=-\frac{e^2}{\hbar}\sum_a\int_{\bm{k}}\epsilon_{ijk}\Omega_{a,\bm{k}}^kf\left(E_{a,\bm{k}}\right)d\bm{k},
\end{align}
where $\epsilon_{ijk}$ is the Levi-Civita symbol, $f\left(E_{a,\bm{k}}\right)$ is the Fermi Dirac distribution function, and the superscript in $\Omega^k_{a,\bm{k}}$ denotes the spatial component of $\bm{\Omega}_{a,\bm{k}}$. Comparing Eq. \ref{OM_edge} and Eq. \ref{J_perp}, we find that the Berry curvature acts as the central hub connecting the orbital magnetization and the Hall conductivity.

The above derivations clearly demonstrate that the linear Hall conductivity arises from the coupling between orbital magnetization and an external electric field. In the linear Hall effect that requires broken time reversal symmetry (TRS), the orbital magnetization in Eq. \ref{OM_edge} is a ground state property that reflects the edge currents circulating around the sample. As can be seen in Fig. \ref{fig1} (a), applying an electric field in the sample plane unbalances the edge currents at opposite edges, so the electrons in opposite propagating channels have different chemical potentials~\cite{Halperin, Yoshioka}. When electrodes are attached to measure the voltage transverse to the edge current direction, the chemical potential difference at opposite edges gives the Hall voltage. The scenario of orbital magnetization with the associated edge currents not only applies well to interpreting the linear Hall effect, but also inspires an insightful understanding of the second order nonlinear Hall effect as presented in the below.

\emph{Second order Hall conductivity derived from orbital magnetization}.--- At linear order, the prerequisite for obtaining a nonzero Hall conductivity in Eq. \ref{Hall01} is to have a finite orbital magnetization in the material. In the linear Hall effect, the finite orbital magnetization can result from either an external magnetic field or a spontaneous TRS breaking. Apart from them, orbital magnetization can also be induced by applying an electric field, which is known as the orbital magnetoelectric effect~\cite{Moore, Wenyu02}. In a material that exhibits active orbital magnetoelectric response, it is conceivable that the orbital magnetization induced by the first order electric field can be further perturbed by the second order electric field. Following the mechanism of Hall voltage generated from the orbital magnetization in the linear Hall effect, we demonstrate that the second order Hall conductivity can be analogously derived from the linear orbital magnetoelectric susceptibility.
\begin{figure}[t]
    \centering
    \includegraphics[width=0.48\textwidth]{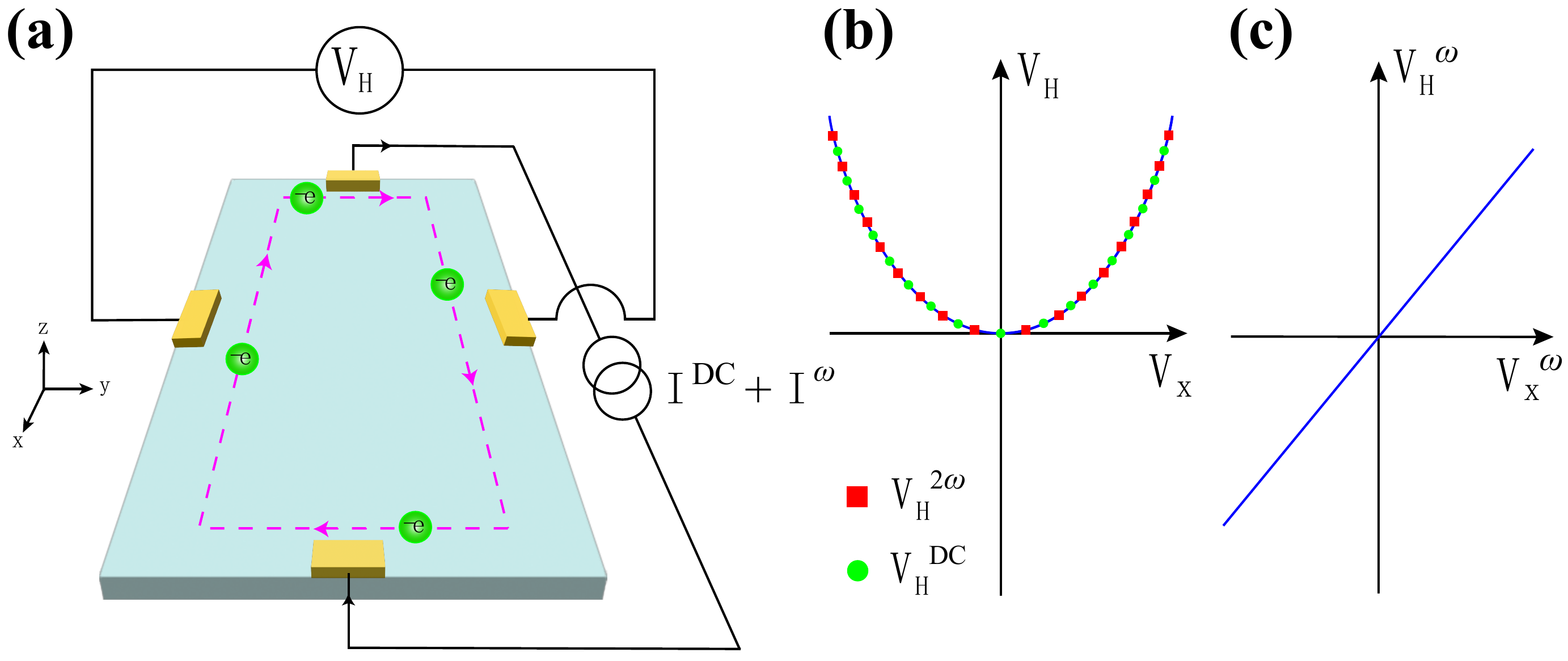}
    \caption{The transport experiment to verify orbital magnetization as the origin of the nonlinear Hall effect. (a) The experimental setup that combines both DC and AC currents. The DC current $I^{\textrm{DC}}$ is to induce the nonlinear Hall effect and the associated orbital magnetization. The AC current $I^\omega\ll I^{\textrm{DC}}$ serves as a perturbation to detect the orbital magnetization. (b) The Hall voltages scale quadratically with the $x$-directional voltage, which confirms the nonlinear Hall effect. (c) The first harmonic Hall voltage scales linearly with the $x$-directional AC voltage, which verifies the orbital magnetization and the affliated edge current.} 
    \label{fig2}
\end{figure}

Applying the perturbation theory to the orbital magnetization in Eq. \ref{OM_edge}, we obtain the electric field induced orbital magnetization $M_i=\chi_{ij}E_j$. Here we have assumed $\bm{M}\left(\bm{E}=\bm{0}\right)=\bm{0}$. Since local magnetic moments contribute negligibly to the macroscopic measurable Hall current~\cite{Supp}, we consider only the orbital magnetization $\bm{M}$ that stems from the edge current. The corresponding orbital magnetoelectric susceptibility is derived to have the form~\cite{Supp}
\begin{align}\label{chi_M}
\chi_{ij}=&-\frac{e^2}{\hbar}\sum_a\int_{\bm{k}}\left(\frac{\tau}{\hbar}\Omega^i_{a,\bm{k}}v^j_{a,\bm{k}}+2\epsilon_{ikq}G^{jk}_{a,\bm{k}}v^q_{a,\bm{k}}\right)f\left(E_{a,\bm{k}}\right)
\end{align}
with $\tau$ being the relaxation time, $\bm{v}_{a,\bm{k}}=\partial_{\bm{k}}E_{a,\bm{k}}$ and $G^{jk}_{a,\bm{k}}$ being the band resolved quantum metric tensor~\cite{Supp, KamalDas,YuandongWang001}. Similar to the linear Hall effect, the electric field induced nonequilibrium orbital magnetization $M_i=\chi_{ij}E_j$ is also accompanied by a Hall current. Substituting $M_i=\chi_{ij}E_j$ and Eq. \ref{chi_M} back to Eq. \ref{J_perp}, we obtain the Hall current density $J_{H,i}=\sigma^H_{ijk}E_jE_k$, where the second order Hall conductivity takes the form
\begin{align}\nonumber\label{2nd_Hall_sigma}
\sigma^H_{ijk}=&-\frac{e^3}{\hbar}\sum_a\int_{\bm{k}}\left[\frac{\tau}{2\hbar}\left(\epsilon_{iqk}\partial_{k_j}\Omega^q_{a,\bm{k}}+\epsilon_{iqj}\partial_{k_k}\Omega^q_{a,\bm{k}}\right)\right.\\
&\left.+2\partial_{k_i}G^{jk}_{a,\bm{k}}-\partial_{k_j}G^{ik}_{a,\bm{k}}-\partial_{k_k}G^{ij}_{a,\bm{k}}\right]f\left(E_{a,\bm{k}}\right).
\end{align}
It is clear that $\sigma^H_{ijk}$ in Eq. \ref{2nd_Hall_sigma} is antisymmetric when exchanging $i$ with $j$ and $k$, consistent with the transverse nature of the Hall response~\cite{Wenyu01,Sirkin00}. Crucially, $\sigma^H_{ijk}$ derived in Eq. \ref{2nd_Hall_sigma} explicitly captures both the intrinsic Berry curvature dipole~\cite{Intisodeman} and quantum metric dipole~\cite{YangGao01, DiXiao, Shengyuan,Kalan} contributions.

\emph{Proposal for the experimental verification}.--- The closely related Eq. \ref{chi_M} and Eq. \ref{2nd_Hall_sigma} demonstrate that the generation of second order Hall voltage response undergoes a nonequilibrium process. In the process, the first order effect of the applied electric field is to induce a nonequilibrium orbital magnetization associated with an edge current, as is shown in Fig. \ref{fig1} (b). Then the second order electric field introduces spatial inhomogeneity to the chemical potentials of the states at opposite edges, generating a second order transverse Hall voltage. In order to verify the above scenario, two critical steps are required: 1) to confirm the affliation between the orbital magnetoelectric effect and the nonlinear Hall effect; 2) to identify the emerging edge states in the current induced orbital magnetization shown in Fig. \ref{fig1} (b).

In a setup illustrated in Fig. \ref{fig2} (a), given a layered material that exhibits a finite nonlinear Hall response, applying a DC current $I^{\textrm{DC}}$ simultaneously generates a quadratic Hall voltage $V_H\propto\left(V^{\textrm{DC}}_x\right)^2$ (Fig. \ref{fig2} (b)) and an orbital magnetization. Here $V^{\textrm{DC}}_x$ denotes the $x-$directional DC voltage. The current indiced orbital magnetization can be identified via either the magneto-optical Kerr rotation microscopy~\cite{JieunLee, JieunLee2} or a scanning superconducting quantun interference device (SQUID)~\cite{AFYoung}. Ideally, a scanning SQUID with sufficiently high spatial resolution can image the edge states in the current induced orbital magnetization~\cite{AFYoung, Moler01, Moler02}. Apart from the SQUID, applying a combination of DC current $I^{\textrm{DC}}$ and AC current $I^{\omega}$ under the condition $I^{\omega}\ll I^{\textrm{DC}}$ provides an alternative way to probe the edge current shown in Fig. \ref{fig1} (b). The AC current $I^\omega$ serves as a perturbation to the orbital magnetization induced by $I^{\textrm{DC}}$ and is expected to generate both the first and second harmonic Hall voltage: $V_H^\omega\propto V_x^\omega$ (Fig. \ref{fig2} (c)) and $V_H^{2\omega}\propto\left(V_x^\omega\right)^2$ (Fig. \ref{fig2} (b)), with $V_x^\omega$ denoting the $x-$directional AC voltage. Here $V_H^{2\omega}$ is the second harmonic component of the nonlinear Hall voltage. Importantly, the Hall voltage $V_H^\omega$ manifests the linear Hall response from the current induced orbital magnetization, so the well established edge current mechanism of the linear Hall effect~\cite{Halperin, Yoshioka} indicates that the observation of $V_H^\omega$ will directly evidence that the current induced orbital magnetization hosts an edge current.

The proposed transport measurements with $I^{\textrm{DC}}+I^\omega$ takes the dual-frequency approach to label the distinct field roles: orbital magnetization generation (by DC) and perturbation (by AC). In most cases, the nonlinear Hall effect does not require the applied electric field to have components of different frequencies. However, in isotropic chiral metals of T and O point groups, electric field components of different frequencies become essential for the nonlinear Hall effect, which further reveals the orbital magnetization origin of the nonlinear Hall effect.

\begin{figure}[t]
    \centering
    \includegraphics[width=0.48\textwidth]{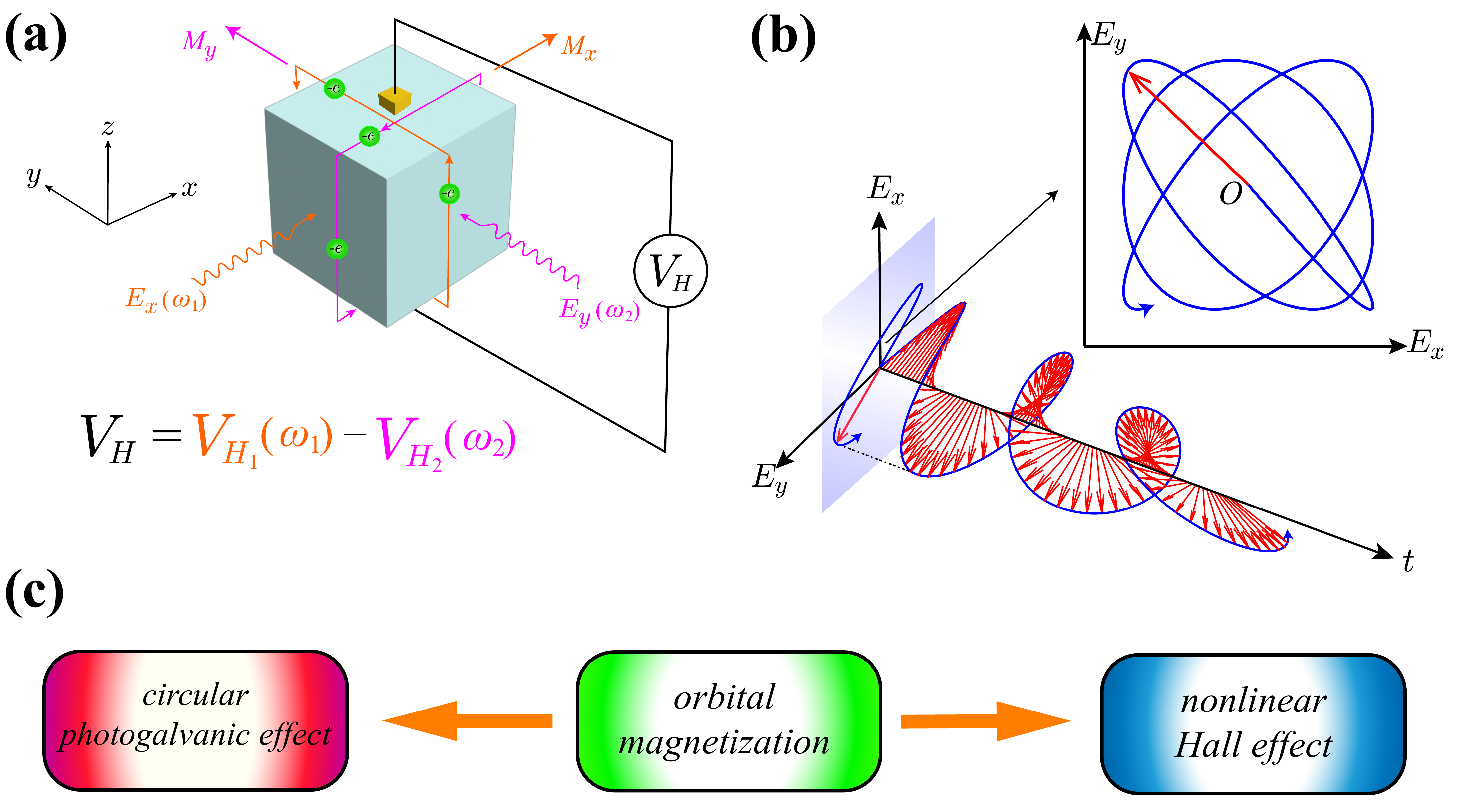}
    \caption{Schematic showing of the nonlinear Hall effect driven by a non-collinear bichromatic electric field. (a) In a chiral metal, the field components $E_x\left(\omega_1\right)$ and $E_y\left(\omega_2\right)$ induce orbital magnetizations along the $x$ and $y$ directions, respectively. The affliated edge currents (the orange and purple loops) are further perturbed by the electric field components, generating $z$-directional Hall voltages $V_{H_1}\left(\omega_1\right)$ and $-V_{H_2}\left(\omega_2\right)$, respectively. The total Hall voltage thus becomes $V_H=V_{H_1}\left(\omega_1\right)-V_{H_2}\left(\omega_2\right)$. (b) The Lissajous curve of a non-collinear bichromatic electric field. The total electric field is represented by red arrows, and its time evolution gives the Lissajous curve colored in blue. The precession of the non-collinear bichromatic electric field generates the dynamical chirality, which drives a fintie nonlinear Hall response in isotropic chiral metals. (c) In isotropic chiral metals, both the circular photo-galvanic effect and the nonlinear Hall effect originate from orbital magnetization.}
    \label{fig3}
\end{figure}

\emph{Nonlinear Hall effect in chiral metals}.--- Isotropic chiral metals of T and O point groups can exhibit a longitudinal orbital magnetoelectric effect with the orbital magnetization direction parallel to the applied current~\cite{Wenyu03, Wenyu04}. While this longitudinal orbital magnetoelectric effect implies a possible nonlinear Hall effect, the symmetry of T and O point groups forces a complete suppression of the nonlinear Hall response if the driving electric field has a single constant frequency~\cite{Wenyu01, Nandy}. This suppression mechanism is schematically illustrated in Fig. \ref{fig3} (a). For a monochromatic electric field $\bm{E}\left(\omega\right)$ in the $xy$ plane, the $x$-component field $E_x\left(\omega\right)$ and $y$-component field $E_y\left(\omega\right)$ induce orbital magnetizations $M_x$ and $M_y$, respectively, with associated edge currents. In principle, $E_x\left(\omega\right)$ perturbs the edge current of $M_y$ to produce a $z$-directional Hall voltage $-V_{H_2}\left(\omega\right)$, while $E_y\left(\omega\right)$ perturbs the edge current of $M_x$ to produce a $z$-directional Hall voltage $V_{H_1}\left(\omega\right)$. However, since $E_x\left(\omega\right)$ and $E_y\left(\omega\right)$ are identical in frequency and phase, the resulting $z$-directionial Hall voltages, $V_{H_1}\left(\omega\right)$ and $-V_{H_2}\left(\omega\right)$, are of opposite sign and interfere destructively, yielding a net zero Hall voltage along the $z$-direction.

It is important to note that this precise cancellation of $V_{H_1}$ and $-V_{H_2}$ requires the two electric field componenets, $E_x$ and $E_y$, to have exactly the same frequency and phase, while any deviation from this condition prevents the cancellation. Therefore, a finite nonlinear Hall current can arise in an isotropic chiral metal when the two perpendicular electric field components, $E_x\left(\omega_1\right)$ and $E_y\left(\omega_2\right)$, have different frequencies and phases. Here, we refer the total electric field comprised of $E_x\left(\omega_1\right)$ and $E_y\left(\omega_2\right)$ with $\omega_1\neq\omega_2$ as a non-collinear bichromatic electric field. Incorporating frequency dependence into the Berry curvature dipole term in Eq. \ref{2nd_Hall_sigma} yields~\cite{Supp}
\begin{widetext}
\begin{align}\label{bichromatic_J}
\bm{J}_H\left(\omega\right)=&\frac{e^3\tau^2\gamma}{2\hbar^2}\int_{-\infty}^\infty\frac{d\omega_1}{2\pi}\int_{-\infty}^\infty\frac{d\omega_2}{2\pi}\frac{\omega_1-\omega_2}{\left(1-i\omega_1\tau\right)\left(1-i\omega_2\tau\right)}i\bm{E}\left(\omega_1\right)\times\bm{E}\left(\omega_2\right)2\pi\delta\left(\omega-\omega_1-\omega_2\right).
\end{align}
\end{widetext}
Here the isotropic chiral crystalline symmetry fixes the Berry curvature dipole as $-\int_{\bm{k}}\sum_a\partial_{k_j}\Omega^i_{a,\bm{k}}f\left(E_{a,\bm{k}}\right)=\gamma\delta_{ij}$ while the quantum metric dipole vanishes by TRS. Consistent with the symmetry constraint, $\bm{J}_H\left(\omega_1+\omega_2\right)\rightarrow\bm{0}$ when $\omega_1\rightarrow\omega_2$.

It is worth noting that the Hall current density $\bm{J}_H\left(\omega_1+\omega_2\right)$ in Eq. \ref{bichromatic_J} is driven by the dynamical chirality $i\bm{E}\left(\omega_1\right)\times\bm{E}\left(\omega_2\right)$~\cite{Smirnova}, with the sign of $\gamma$ determined by the crystal chirality~\cite{HongDing}. In Fig. \ref{fig3} (b), the Lissajous curve of the non-collinear bichromatic electric field shows that the field oscillates and rotates simultaneously, resembling the precession of a circularly polarized electric field~\cite{Supp} and leading to the dynamical chirality. As $\omega_1\rightarrow-\omega_2$, the dynamical chirality $i\bm{E}\left(\omega_1\right)\times\bm{E}\left(\omega_2\right)$ approaches the static circular polarizability, reducing Eq. \ref{bichromatic_J} to the circular photo-galvanic current density~\cite{Orenstein}. Thus, the circular photo-galvanic effect is a special case of the nonlinear Hall effect and also originates from orbital magnetization (see Fig. \ref{fig3} (c)).

\begin{figure}[t]
    \centering
    \includegraphics[width=0.48\textwidth]{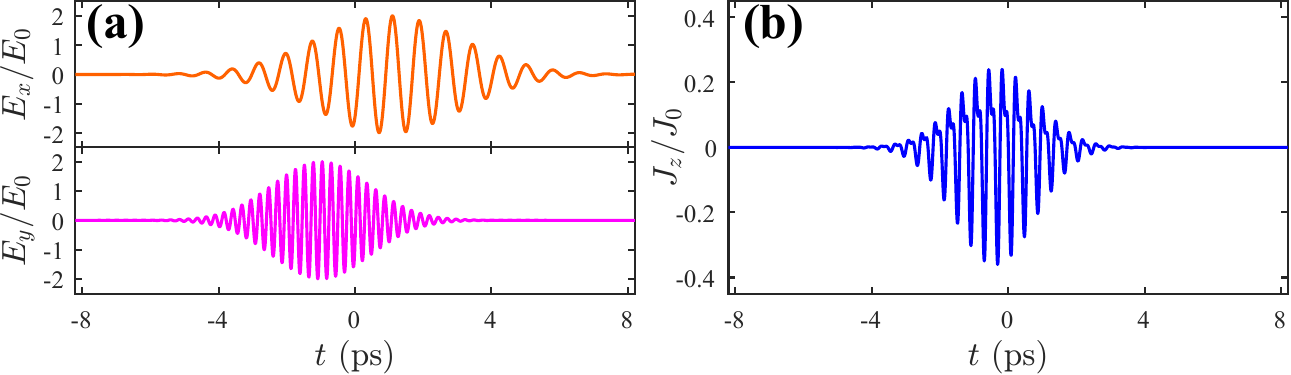}
    \caption{The nonlinear Hall current density driven by a Gaussian pulse field $\bm{E}\left(t\right)=\left[E_1\left(t-\frac{\tau_0}{2}\right),E_2\left(t+\frac{\tau_0}{2}\right),0\right]$ with $E_i\left(t\right)=2E_0\sin\left(\omega_it+\varphi_i\right)\exp\left(-\frac{t^2}{4\sigma_i^2}\right)$, $i=1,2$. (a) The $x$ and $y$ components of the Gaussian pulse field $\bm{E}\left(t\right)$ in the time domain. Here in the simulation, we have set $\omega_1=8~\textrm{THz},\omega_2=24~\textrm{THz},\sigma_1=1.5~\textrm{ps},\sigma_2=1.0~\textrm{ps},\varphi_1=\frac{\pi}{4},\varphi_2=\frac{\pi}{5},\tau_0=2.0~\textrm{ps}$, so the Gaussian pulses in the two perpendicular directions are not synchronized. (b) The resulting Hall current pulse $\bm J_H(t)$ in the $z$ direction. Here, $J_0=\frac{e^3\gamma E_0^2}{\hbar^2}$.} 
    \label{fig4}
\end{figure}

To further clarify the sequential roles of the electric field components in generating nonlinear Hall current, we obtain the time domain Hall current from Eq. \ref{bichromatic_J} to be~\cite{Supp}
\begin{align}\label{Hall_current}
\bm J_H(t)=\frac{e^3\gamma}{\hbar^2}\int^{\infty}_{-\infty}\Theta(t-t')\bm E(t')\times\bm E(t) dt',
\end{align}
where the Heaviside step function $\Theta\left(t-t'\right)$ enforces the causal sequencing of the electric field components. For Gaussian pulses $\bm{E}\left(t\right)=\left[E_1\left(t-\frac{\tau_0}{2}\right),E_2\left(t+\frac{\tau_0}{2}\right),0\right]$ with $E_i\left(t\right)=2E_0\sin\left(\omega_it+\varphi_i\right)\exp\left(-\frac{t^2}{4\sigma_i^2}\right)$ and $i=1,2$, simulations in Fig. \ref{fig4} and Supplemental Materials~\cite{Supp} show that $\bm{J}_H(t)$ is always finite as long as the two components of $\bm{E}\left(t\right)$ are not synchronized, consistent with the sequential roles of the electric field required for the nonlinear Hall effect. In fact, Eq. \ref{Hall_current} demonstrates that the chiral correlator $\bm{\mathcal{C}}\left(t,t'\right)=\Theta\left(t-t'\right)\bm{E}\left(t'\right)\times\bm{E}\left(t\right)$ acts as the kernal to generate $\bm{J}_H\left(t\right)$~\cite{Supp}, which means that the nonlinear Hall current in isotropic chiral metals directly manifests the chiral correlation of the applied electric field.

Among the 18 gyrotropic point groups that allow finite orbital magnetoelectric effect~\cite{Wenyu03}, the T and O point groups are unique in requiring non-collinear bichromatic electric fields for nonlinear Hall effects. Recently, a series of B20 transition metal monosillicides (e.g., RhSi, CoSi)~\cite{Burkov}, which are isotropic chiral metals of T point group, have been identified to host multi-fold Weyl points. As Weyl points act as source of Berry curvature in $\bm{k}$ space, these materials are considered as prime candidate to exhibit nonlinear Hall effects. For RhSi, its Berry curvature dipole reaches $\gamma=0.04$ near the Fermi energy (Fig. S2), comparable to that in WTe$_2$~\cite{YangZhang02}. Given an estimated relaxation time in the order of 1fs~\cite{Rees, LiangWu2}, the second order Hall conductivity can reach the order of $\mu$A/V$^2$ under driving fields of Terahertz. As a giant circular photo-galvanic effect has been observed in RhSi~\cite{Rees, LiangWu2}, we expect that the related non-collinear bichromatic electric field driven nonlinear Hall effect is also observable in RhSi.

\emph{Discussions}.--- In this work, we establish the orbital magnetization as the origin of the nonlinear Hall effect, which mainly applies to the intra-band process. A further generalization of our scenario is to take into account inter-band processes that bypass the constraint of Fermi surfaces, so insulators are also expected to exhibit a finite nonlinear Hall response~\cite{Wenyu01}. As the orbital magnetization always occurs in quantum Hall insulators, an extension of our scenario of the second order Hall response to the third order one is expected to provide a possible interpretation to the recently observed third order nonlinear Hall effect in the graphene quantum Hall system~\cite{PanHe}. It is noteworthy that our current scenario only deals with the intrinsic band contributions to the nonlinear Hall effect. For the disorder scattering induced nonlinear Hall phenomenon~\cite{Haizhou2, ZZDu03}, its relation to the orbital magnetization needs further exploration.

Starting from the scenario of orbital magnetization origin, we have pointed out that isotropic chiral metals of T and O point groups host a unique nonlinear Hall effect that is driven by a non-collinear bichromatic electric field. Such nonlinear Hall effect is found to manifest the chiral correlation of the applied electric field, while the Hall response vanishes if the field components are in-phase and monochromatic. This lays the foundation for the chiral coherent spectroscopy~\cite{Hache, YRShen}, where applying a non-collinear bichromatic field and measuring the sum-frequency generation in the transverse direction can identify the chiral structure of crystals. In the other way around, since isotropic chiral crystals can enable the circular polarized part of the applied electric field to exhibit a transverse sum-frequency generation, one can utilize isotropic chiral crystals to quantify the chiral correlation of the applied electric field.

\emph{Acknowledgements}.--- W.-Y.H. acknowledges the support from the National Natural Science Foundation of China (No. 12304200), the BHYJRC Program from the Ministry of Education of China (No. SPST-RC-10), the Shanghai Rising-Star Program (24QA2705400), and the start-up funding from ShanghaiTech University.

\bibliography{main}

\end{document}